\documentclass[reprint,amsmath,amssymb,aps,prl]{revtex4-2}
\usepackage{graphicx}
\usepackage{dcolumn}
\usepackage{bm}
\usepackage{physics}
\usepackage{xcolor}
\usepackage{pdfpages}

\makeatletter
\newcommand*{\rom}[1]{\expandafter\@slowromancap\romannumeral #1@}
\makeatother

\makeatletter
\AtBeginDocument{\let\LS@rot\@undefined}
\makeatother

\begin{document}

\title{Conservation of angular momentum on a single-photon level}

\author{L. Kopf$^1$}
\email{lea.kopf@tuni.fi}
\author{R. Barros$^1$}%
\author{S. Prabhakar$^2$}%
\author{E. Giese$^3$}%
\author{R. Fickler$^1$}%
\affiliation{$^1$ Tampere University, Photonics Laboratory, Physics Unit, Tampere, FI-33014, Finland}
\affiliation{$^2$ Quantum Science and Technology Laboratory, Physical Research Laboratory, Ahmedabad, India 380009
}%
\affiliation{$^3$ Technische Universität Darmstadt, Fachbereich Physik, Institut für Angewandte Physik, Schlossgartenstr. 7, 64289 Darmstadt, Germany}%

\date{\today}

\begin{abstract}
Identifying conservation laws is central to every subfield of physics, as they illuminate the underlying symmetries and fundamental principles.
A prime example can be found in quantum optics:
The conservation of orbital angular momentum (OAM) during spontaneous parametric down-conversion (SPDC) enables the generation of a photon pair with entangled OAM.
In this article, we report on the first study of OAM conservation in SPDC pumped by single photons.
Our results present the first implementation of cascaded down-conversion without waveguides, setting the stage for experiments on the direct generation of multi-photon high-dimensional entanglement using all degrees of freedom of light.
\end{abstract}

\keywords{Cascaded SPDC, down-conversion, momentum conservation, OAM}
\maketitle

Symmetries and conservation laws are at the heart of our natural scientific understanding of the world, as they explain why certain phenomena do or do not exist in nature.
In the realm of classical nonlinear optics---where light-matter interactions enable the conversion of light to different wavelengths---these principles are also essential \cite{boyd2008nonlinear,walborn2010spatial}:
The conservation of energy dictates allowed wavelength combinations. 
Symmetries in the crystalline structure constrain the possible polarizations of the interacting fields. 
The conservation of linear momentum is conditioned on the translational symmetry of the medium.
In the longitudinal direction, momentum conservation gives rise to the phase-matching conditions.
In the transverse plane, momentum conservation defines a set of selection rules for the spatial structures of the interacting waves.
One particularly prominent example of such selection rules is the conservation of orbital angular momentum (OAM), which originates from the beams' transverse spatial structure~\cite{Simon2021Topology}.
The prime example of a light field carrying OAM is an optical vortex, defined by an azimuthal phase gradient $\exp({\text{i}\ell\phi})$ along the angle $\phi$, where $\ell$ is called the topological charge and defines the OAM of the light field~\cite{Allen1992Orbital,he1995direct}.
Subsequent studies revealed that nonlinear optical processes, such as second harmonic generation, conserve the OAM of interacting fields \cite{dholakia1996second}.
While initially challenged in the context of non-perturbative high-harmonic generation \cite{zurch2012strong}, OAM conservation has been confirmed for such processes \cite{hernandez2013attosecond,gariepy2014creating,geneaux2016synthesis}, establishing it as the standard intuitive explanation for the frequency conversion of light carrying OAM.

Since connected to the properties of modes, this concept can be transferred to the quantum regime where it plays a crucial role in the interpretation of quantum operations, in particular spontaneous parametric down-conversion (SPDC).
Commonly, SPDC is implemented with a second-order nonlinear medium illuminated by a classical pump field, i.\,e., a strong coherent laser, resulting in the spontaneous emission of a signal and idler photon pair.
Correlations between the emitted photons have since become a standard tool to either realize a source of heralded single photons or to generate photonic entanglement between the two (or more) photons \cite{pan2012multiphoton, erhard2020advances}. 

The spatial correlations of the photon pairs generated in SPDC have been subject to many recent studies, ranging from the realisation of the Einstein-Podolski-Rosen paradox via position-momentum entanglement ~\cite{EPR_Original,Boyd_EPR} to quantum imaging \cite{lemos2014quantum,padgett2017introduction}.
Despite an early SPDC experiment that seemed to violate the conservation of OAM \cite{Arlt1999Parametric}, the seminal experiment by Mair et al. \cite{Mair2001Entanglement} has shown that SPDC driven by a laser field conserves OAM \cite{osorio2008correlations, feng2008spatial}.
These spatial quantum correlations have profound implications for fundamental quantum science and modern quantum technologies, as shown by a plethora of studies \cite{arnaut2000orbital, franke2002two, molina2007twisted, leach2010quantum, krenn2017orbital, srivastav2022characterizing}.

OAM conservation in SPDC is intuitively understood as the transfer of the topological charge of a photon from the strong coherent pump field to a photon pair.
Since the OAM of a light beam is given by $\hbar \ell$ per photon, the photon-number fluctuations inevitably associated with a strong coherent pump field also imply fluctuations of the overall OAM. 
Hence, SPDC experiments using a classical pump can only witness the conservation of the average OAM in the pump field.
In this scenario, a natural question arises:
How is it possible to confirm that each quantum of OAM is in fact conserved in the SPDC process?

In revisiting the symmetries of SPDC and the correlation of the associated modes carrying OAM, we confirm that the conservation of OAM indeed holds on the single-photon level and not only for an ensemble average. 
Despite the extremely low count rates that result from the necessity of using bulk nonlinear crystals with low conversion efficiencies instead of waveguides, we present experiments that verify the conservation of OAM for pump photons with up to two quanta of OAM.
To further support our results, we compare measured OAM correlations with those obtained using a classical pump field and confirm that there is no measurable difference.
Finally, we present indications that the generated photons are entangled.
Hence, our work not only demonstrates OAM conservation at the fundamental quantum level but, owing to the use of bulk crystals instead of waveguides, also paves the way for heralded spatially-entangled photon pair sources and high-dimensionally entangled three-partite photonic states invoking all degrees of freedom of light.

\vspace{0.5em}
\noindent\textit{Fundamentals.}\textemdash 
In SPDC, an input pump field ($p$) in state $|\psi_p\rangle$ is down-converted into signal ($s$) and idler ($i$) photons.
Neglecting the tensor nature of the crystal and by that walk-off effects, the nonlinear interaction is described by a time-dependent Hamiltonian~\cite{walborn2010spatial,karan2020phase}
\begin{equation}
    \hat{H}\! \propto \!\! \int \text{d}^3\boldsymbol{r}\, \chi^{(2)}(z) \hat{E}_p^{(+)}(\boldsymbol{r},t)\hat{E}_s^{(+)\dagger}(\boldsymbol{r},t)\hat{E}_i^{(+)\dagger}(\boldsymbol{r},t) + \textrm{h.c.}\,,
    \label{Hamiltonian}
\end{equation}
where $\text{h.c.}$ denotes the Hermitian conjugate and $\chi^{(2)}$ the spatially dependent nonlinear susceptibility of the medium.
The positive-frequency part of the electric field operator of field $j=p,s,i$ is given by
\begin{equation}
    \hat{E}_j^{(+)}(\boldsymbol{r},t) \propto \sum_{p_j\ell_j} u_{p_j\ell_j}(\boldsymbol{r}) e^{i[k_j(\omega_j)z -\omega_j t]}\hat{a}_{p_j\ell_j} \, .
    \label{FieldOperator}
\end{equation}
Here, we chose a decomposition into Laguerre-Gaussian (LG) paraxial modes $u_{p_j\ell_j}(\boldsymbol{r})=~u_{p_j\ell_j}(\boldsymbol{\varrho},z)$ that depend on the transverse coordinates $\boldsymbol{\varrho}$ and propagate in the $z$ direction with a central wave number $k_j(\omega_j)$, associated with the frequency $\omega_j$.
The bosonic operator $\hat{a}_{p_j\ell_j}$ annihilates a photon in the mode associated with the indices $p_j$ and $\ell_j$ that refer to the mode's radial order and the topological charge, respectively.
These operators fulfill the bosonic commutation relations $[\hat{a}_{p_j\ell_j},\hat{a}_{p^\prime_{j^\prime}\ell^\prime_{j^\prime}}^\dagger]=\delta_{p_{j}^{\textcolor{white}{\prime}},p^\prime_{j^\prime}} \delta_{\ell_j^{\textcolor{white}{\prime}},\ell_{j^\prime}^\prime}$.

The interaction Hamiltonian is now expressed by
\begin{align}
\hat{H}\propto \!\sum\limits_{\{p_j, \ell_j\}} \text{e}^{- \text{i}\Delta \omega t}  \Lambda_{\{\ell_j\}}^{\{p_j\}}\hat{a}_{p_p\ell_p}\hat{a}^\dagger_{p_s\ell_s}\hat{a}^\dagger_{p_i\ell_i} + \textrm{h.c.}\,,
\label{SimplifiedHamiltonian}
\end{align}
where $\Delta\omega =  \omega_p-\omega_s - \omega_i$ is the frequency mismatch of the interacting fields.
The summation is taken over all mode indices $\{p_j, \ell_j\}$.
The transverse mode overlap integral is given by
\begin{eqnarray}
\Lambda_{\{\ell_j\}}^{\{p_j\}} = \!\! \int \!\! \text{d}^3\boldsymbol{r}~ \chi^{(2)}(z)\,e^{i\Delta kz} u_{p_p\ell_p}(\boldsymbol{r}) u^*_{p_s\ell_s}(\boldsymbol{r}) u^*_{p_i\ell_i}(\boldsymbol{r}) , 
\label{OverlapIntegral}
\end{eqnarray}
where $\Delta k = k_p(\omega_p) -k_s(\omega_s)-k_i(\omega_i)$ is the wave vector mismatch.
It represents the inner product between the pump mode and the product of the signal and idler modes. 
The overlap integral is a classical quantity describing the spatial modes used in the canonical quantization procedure, and an identical expression appears in the classical nonlinear optical theory with structured light fields \cite{schwob1998transverse, alves2018conditions}.
The integral over the longitudinal direction is taken over the length of the crystal and accounts for the phase mismatch between the interacting waves, while the overlap over  the transverse coordinates gives rise to the spatial selection rules.
Due to the rotational symmetry of the LG modes we observe in cylindrical coordinates $u_{p_j\ell_j}(\varrho,\phi,z)\propto \exp (\text{i} \ell_j \phi)$ and therefore $\Lambda_{\{\ell_j\}}^{\{p_j\}} \propto \int_0^{2\pi} \text{d}\phi \exp [\text{i}(\ell_p-\ell_s-\ell_i)\phi]\propto \delta_{\ell_p,\ell_s+\ell_i}$, i.\,e., the symmetry of the classical modes defines the selection rules of nonlinear optics which include---but are not limited to---the conservation of the topological charge. 
While in fact the rotational symmetry is formally broken due to birefringence, such effects can be neglected for our experimental configuration~\cite{Martinelli2004}, justifying the form of the Hamiltonian in Eq.~\eqref{Hamiltonian}.
Note, that such a symmetry does not apply to the radial index, and no analogue conservation property for the indices $p_j$ exists \cite{DErrico2021Full,miatto2011full}.

The above discussion demonstrates that the conservation of OAM in SPDC is caused by a symmetry property of classical modes that dictates which modes couple.
Of course, it also has implications for quantum optics operating in the spontaneous regime.
Usually, correlations of the generated signal and idler photons in their respective OAM modes are measured. 
To study the conservation of OAM, we use the photon number operator $\hat{n}_{p_j\ell_j} =\hat{a}_{p_j\ell_j}^\dagger\hat{a}_{p_j\ell_j}$ to define an OAM operator $\hat{L}_\text{OAM} =\hbar \sum_j\sum_{p_j,\ell_j}\ell_j\,\hat{n}_{p_j\ell_j}$ ~\cite{grynberg2010introduction}.
Since $[\hat{L}_\text{OAM}, \hat{a}_{p_p\ell_p} \hat{a}^\dagger_{p_s\ell_s}\hat{a}^\dagger_{p_i\ell_i}]= \hbar (\ell_s+\ell_i - \ell_p) \hat{a}_{p_p\ell_p} \hat{a}^\dagger_{p_s\ell_s}\hat{a}^\dagger_{p_i\ell_i}$ and $\Lambda_{\{\ell_j\}}^{\{p_j\}} \propto  \delta_{\ell_p,\ell_s+\ell_i}$, we observe a vanishing commutator $[\hat{L}_\text{OAM}, \hat{H}]=0$.
Consequently, the Hamiltonian from Eq.~\eqref{SimplifiedHamiltonian} implies that the OAM operator is conserved. 
Of course, this property transfers both the expectation value $\langle\hat{L}_\text{OAM}\rangle$ and its variance $\Delta L_\text{OAM}^2$, such that OAM conservation holds on average and for all input pump fields.
Although the conservation of OAM has been confirmed using laser fields to drive the process, demonstrating this conservation law for processes induced by quantum states, like single-photon states, has yet to be achieved.
Hence, for a genuine observation of OAM conservation in SPDC on the single-photon level a single-photon Fock state pump must be used.

\begin{figure*}[ht] 
    \centering
    \includegraphics[width =0.85\textwidth]{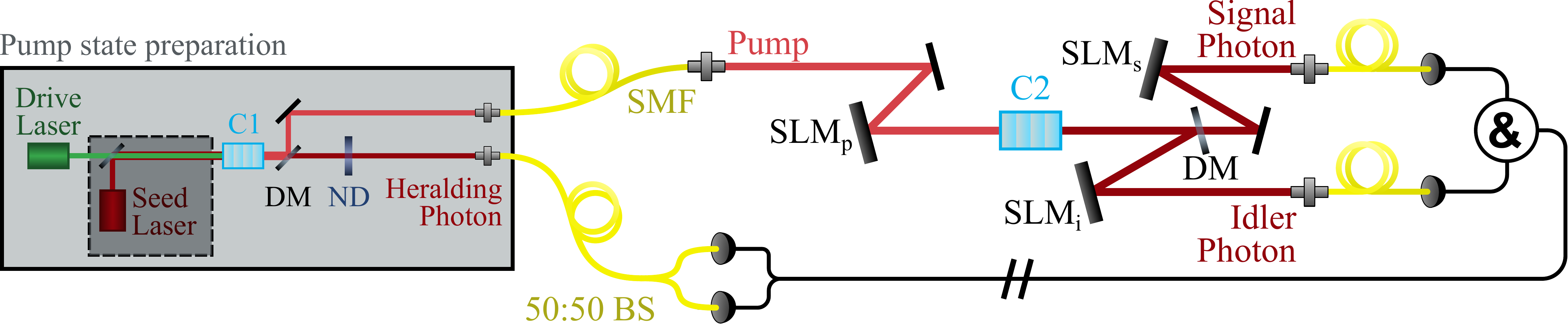}
    \caption{Experimental setup: 
    First, the pump state is prepared in a PDC process in a first nonlinear process (C1).
    The process is switched between spontaneous and stimulated photon emission by using a seed laser, yielding a heralded single-photon state and a weak coherent state, respectively.
    When generating a heralded single-photon the heralding photon is separated from the pump state with a dichroic mirror (DM). The heralding signal is then attenuated with a neutral density (ND) filter and split-up with a 50:50 fiber beam splitter (50:50 BS) to prevent the saturation of the detectors.
    The pump field is collected by a single-mode fiber (SMF), which guarantees a near Gaussian spatial profile, and shaped with a spatial light modulator (SLM) before driving a second nonlinear process (C2).
    The generated signal and idler photons are projected onto different spatial modes by phase-flattening their wavefronts with SLM\textsubscript{s} and SLM\textsubscript{i}, respectively, and collecting them with SMFs.}
    \label{fig1}
\end{figure*}
\vspace{0.5em}
\noindent\textit{Experiment.}\textemdash 
To implement an SPDC experiment with a Fock state pump, we use the setup shown in Fig.~\ref{fig1}.
It consists of two cascaded SPDC sources, where the single photons generated in the first source are used to pump the second source.  
Cascaded SPDC has been demonstrated earlier for the direct generation of three photons \cite{Hübel2010Direct} that can be entangled in polarization \cite{Hamel2014Direct, chaisson2022phase} or time/energy \cite{shalm2013three} to observe genuine three-photon interference \cite{agne2017observation}, and the certification of a polarization qubit \cite{meyer2016certifying}.
These experiments, however, were implemented with efficient waveguides as the second SPDC process, which are restricted to a single spatial mode and thus cannot support photons carrying OAM.
In contrast, the second SPDC process in our experiment is built with bulk optics, which allows OAM-carrying pump, signal, and idler photons.

The pump state is prepared in the first SPDC source, containing a 20\,mm long type-0 periodically poled potassium titanyl phosphate (ppKTP, Raicol) crystal with a poling period of 9.375\,{\textmu}m. 
The source is driven by a continuous wave laser centered around 524\,nm, which we call the drive field, that is temperature-tuned for the generation of highly non-degenerate photon pairs at wavelengths of around 783\,nm and 1588\,nm.
The single photon source has negligible multi-pair emissions as discussed in Supplementary Note \rom{2} \cite{suppl}.
The 783\,nm photon is used to pump the second SPDC source, while the 1588\,nm photon is used to herald the pump photon.
In addition, a seed laser matching the wavelength and spatial mode of the heralding signal is used to stimulate the PDC process, allowing the switch between a heralded single-photon pump state (spontaneous) and a coherent pump state (stimulated).
The seed laser has a power of 12.2\,mW, a bandwidth of around 80\,kHz, and is tuned to have the same polarization as the driving field.
The stimulated process results in a quasi-classical pump \cite{grynberg2010introduction} with 12 {\textmu}W of power, which is also used for aligning the second SPDC source.

After coupling the pump and heralding photons to single-mode fibers (SMFs), we achieve a heralded and unheralded single-photon rate of 464\,kHz and 2.01\,MHz per mW of the drive field, respectively, where the heralding photons are measured with superconducting nanowire single-photon detectors (Single Quantum) with high detection efficiencies ($\approx$ 80\,\%) and low dark counts ($\leq100$\,Hz).
The detection of the high heralding rates pose an experimental challenge. To achieve the maximal heralding rate, which is only limited by the saturation of the heralding detectors, we could (i) reduce the driving power or (ii) introduce loss in the heralding arm. By reducing the driving power, we would also reduce the rate of pump photons and consequently the generation rate of unheralded signal and idler photons. Instead, we introduce losses in the heralding arm with a neutral density filter to reach the saturation limit of the detector, which results in the same maximal heralded signal and idler rate, however, with a much larger rate of unheralded signal and idler photons.
Lastly, the pump mode out-coupled from an SMF is shaped by a spatial light modulator (SLM\textsubscript{p}, HoloEye Pluto) displaying a spiral phase $\exp(\text{i}\ell_p \phi)$, which sets the OAM of the pump to $\ell_p\hbar$ per photon.
No amplitude modulation technique is implemented to minimize losses that come with more advanced shaping techniques \cite{bolduc2013exact}.

The second SPDC source uses a 25\,mm long type-0 periodically poled lithium niobate (ppLN, HC Photonics) crystal with a poling period of 19.40\,{\textmu}m.
The crystal is temperature-tuned to optimize the down-conversion of the 783\,nm pump field into collinear signal and idler photons centered at around 1534\,nm and 1600\,nm, respectively, which can be efficiently separated with a dichroic mirror.
The OAM values of the signal and idler photons are then measured through phase-flattening with an SLM and SMF-coupling \cite{Forbes2016Creation}.

The generated OAM spectrum of the second SPDC source is set by tuning the ratio $R_{w_0}$ between the pump and the signal/idler beam waists at the center of the second nonlinear crystal to optimize the SPDC efficiency for $|\ell_{s,i}|\le1$ for each pump mode.
Using the coherent pump field and setting $\ell_{p,s,i}=0$, which is the configuration with the highest efficiency, we achieve a photon pair generation rate of about 601\,kHz per mW pump power.
For a heralded single-photon pump in the same setting, we achieve a coincidence rate of $1.3\pm0.4$ heralded photon pairs per hour, which is high above the measured accidental rate of 0.1 per hour.
When the heralding event is disregarded, the total rate of signal/idler coincidences is $40.2\pm1.7$ per hour.
The accidental rate is the mean measured coincidence rate outside of the coincidence window, as described in Supplementary Note \rom{1} \cite{suppl}.

\begin{figure}[t] 
    \centering
    \includegraphics[width = 0.325\textwidth]{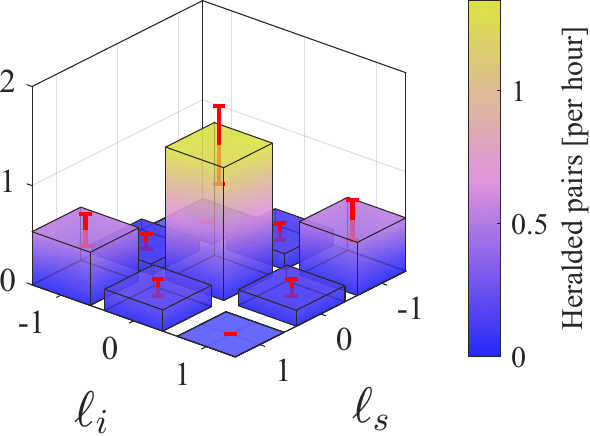}
    \caption{Signal/idler OAM correlation matrices obtained with an $\ell_p=0$ pump mode with a heralded single-photon state.
    Signal and idler photons are projected on 9 different combinations of OAM quanta.
    When OAM conservation applies, only projections on the diagonal of the matrix should be populated ($\ell_s = -\ell_i$).
    The total measurement time is 168.0\,h and the measurements are corrected for accidental coincidences.
    The error bars give the standard deviations for 1.5\,h time bins.
    } 
    \label{fig2}
\end{figure}

\begin{figure*}[ht]  
    \centering
    \includegraphics[width = \textwidth]{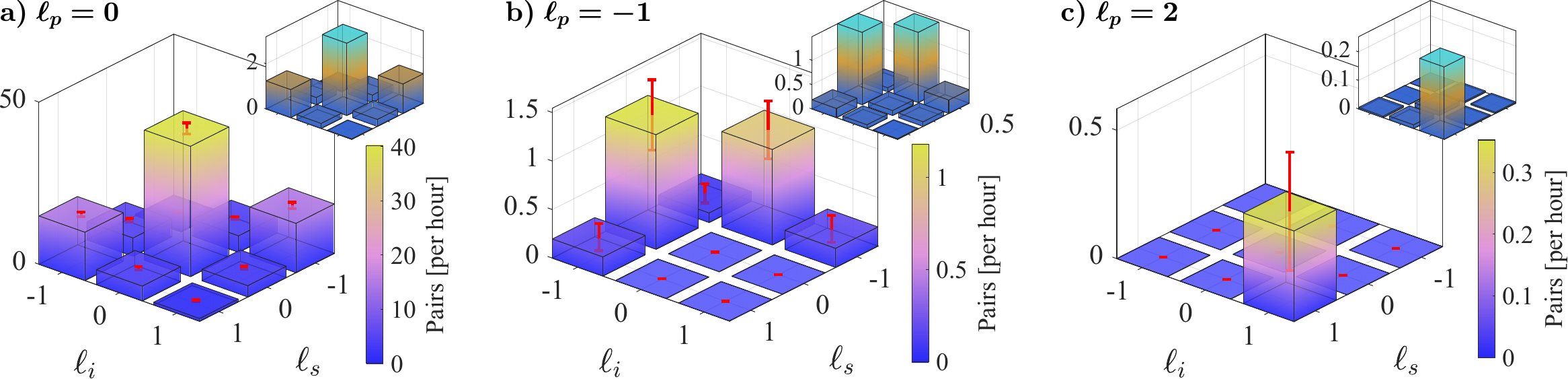}
    \caption{OAM correlation matrices for pump photons carrying different OAM values, namely (a) $\ell_p=0$, (b) $\ell_p=-1$, and (c) $\ell_p=2$, measured over 168.0\,h, 91.5,h, and 76.5\,h, respectively.
    The data is corrected for accidentals, and the errors are standard deviations for 1.5\,h time bins.
    The insets show the corresponding correlation matrices measured with a coherent pump where the coincidence rates are given in kHz, with error bars omitted for clarity.} 
    \label{fig3}
\end{figure*}

\vspace{0.5em}
\noindent\textit{Measurements.}\textemdash  
Figure~\ref{fig2} shows the OAM correlation matrix of signal/idler photon pairs generated with a heralded single-photon pump with no OAM, i.e., $\ell_p=0$ and $R_{w_0}=2.4\,$.
We cyclically measure each combination of $\ell_s$ and $\ell_i$ modes for around 10\,min over a total of 168.0\,h, which results in 112 measurements of the entire correlation matrix and a total of 57 heralded photon pairs.
As expected, we observe a reduced number of detections for OAM-carrying photons due to the optimization for maximal efficiency, i.e., the SPDC is optimized for a minimal number of OAM modes.
More importantly, the coincidence rates in all possible OAM combinations show the conservation of OAM, with around 76\% of the overall detections obeying $\ell_s = \ell_p-\ell_i$ limited by experimental imperfections.
We find a very similar distribution of counts when compared to the classical pump case [shown in Fig.~\ref{fig3}\,(a) in the inset], which is also reflected in a sample Pearson correlation coefficient of $c_P=99.5$\% between the two measurements.

Next, we measure the correlations for a single-photon pump carrying OAM.
However, when imprinting an OAM onto the pump, the count rates significantly drop such that we can only measure correlations with an unheralded single-photon pump.
Even though the unheralded pump photon state is not a pure single-photon state, multi-pair emissions from the first SPDC are too low to be measured in our setup. 
However, we estimate the rate of multi-pair emissions using our conversion efficiencies and pump powers and conclude that in average only one out of 16 SPDC events in the first crystal process are multiple pairs (see supplementary Note \rom{2} \cite{suppl}).
To verify that the heralding does not influence the OAM correlation, in Fig.~\ref{fig3}\,(a) we show the correlation matrix with an unheralded single-photon pump with $\ell_p=0$ (along with the one obtained from a classical pump in the inset).
The similarity of the unheralded and heralded correlation matrices is represented by $c_P=99.5$\%, while the unheralded and coherent pump correlation matrices have $c_P=99.7$\%.

We then imprint OAM values of $\ell_{p}=-1$ and $+2$ onto the single-photon pump, and again measure the OAM correlations of signal and idler photons.
For a single-photon pump with $\ell_p=-1$ and $R_{w_0}=3.3$, we measure count rates of $1.2\pm0.4$ coincidences per hour over a total of 91.5\,h.
In full accordance with OAM conservation, only the projections on $\ell_s=0$ and $\ell_i=-1$, and $\ell_s=-1$ and $\ell_i=0$ contribute to the correlation matrix, as is shown Fig.~\ref{fig3}\,(b). 
Again, there is no significant difference to the measurement with a classical pump [see inset in Fig.~\ref{fig3}\,(b) with $c_P=99.1$\%].
As a final measurement, we increase and change the sign of the OAM value of the pump photons, setting $\ell_p=2$ and $R_{w_0}=4.3$, and measure the correlation matrix for a total time of 76.5\,h.
In agreement with OAM conservation, we only observe counts for states with $\ell_{s,i}=1$ at a rate of $0.4\pm0.2$ per hour [see Fig.~\ref{fig3}\,(c)].
Although the measurements also match the correlations observed with a quasi-classical pump ($c_P=99.9$\%), the statistical errors are non-negligible (see supplementary Note \rom{3} \cite{suppl} for the total number of counts of all measurements).
Nevertheless, all our experiments strongly comply with OAM conservation, thus verifying this fundamental conservation law on a full quantum level.

Moreover, we performed additional correlation measurements in a mutually unbiased basis to reveal OAM entanglement. As we present in Supplementary Note \rom{4}, our results surpass the classically achievable bound for the entanglement witness introduced in \cite{Sprengler2012Entanglement}. 
While promising, these preliminary findings do not constitute a definitive proof of entanglement since the measurements were affected by a substantial degradation in the power of the drive laser, resulting in low statistical significance.
To conclusively verify OAM entanglement in this system, additional experiments with consistent experimental conditions will be necessary.

\noindent\textit{Discussion.}\textemdash  
We present a study to test and verify the fundamental law of OAM conservation in the nonlinear optical process of spontaneous parametric down-conversion on a single-photon level.
Before, only correlations generated with coherent pump states have been studied, where fluctuations in photon numbers translate to a nonzero variance in OAM.
Here we show experimentally that even for a single-photon pump, photon pairs generated through SPDC carry OAM values that follow OAM conservation. 
Despite the low conversion efficiencies of bulk SPDC crystals, we are able to record correlation matrices for pump photons with up to two OAM quanta. 
In the future, we expect our results to be enhanced through nonlinear processes with higher nonlinear conversion efficiencies \cite{leger2023amplification}, detectors tolerating higher heralding count rates, the use of deterministic single-photon sources, or by utilizing more efficient OAM measurement schemes, e.\,g. mode sorters \cite{Berkhout2010Efficient, Mirhosseini2013Efficient, Sahu2018Angular, Fontaine2019Laguerre}.
These technical improvements would enable the study of the conservation of larger OAM quanta, the verification of the heralded generation of high-dimensional OAM entanglement, and the exploration of more complex light structures.
Lastly, note that a weak coherent pump state would also be a good approximation to a single-photon Fock state.
However, our experimental setup with cascaded SPDC is not relying on nonlinear processes in waveguides and, thus, can be leveraged for the direct generation of three-photon entanglement in the spatial domain or heralded OAM-entangled photon pairs. 
Together with the already presented work on polarization \cite{Hamel2014Direct, chaisson2022phase} and time/energy \cite{shalm2013three} entanglement, the scheme will enable the generation of multi-partite entanglement in all degrees of freedom of light. 
\newline

\noindent\textit{Acknowledgements.}\textemdash
The authors thank Robert W. Boyd and Boris Braverman for fruitful discussions at an early stage of the experiment. 
The authors acknowledge the support of the Research Council of Finland through the Photonics Research and Innovation Flagship (PREIN - decision 346511).
LK acknowledges the support of the Vilho, Yrjö and Kalle Väisälä Foundation of the Finnish Academy of Science and Letters.
RB acknowledges the support of the Research Council of Finland through the postdoctoral researcher funding (decision 349120). 
RF acknowledges the support of the Research Council of Finland through the Academy Research Fellowship (decision 332399).

\providecommand{\noopsort}[1]{}\providecommand{\singleletter}[1]{#1}%

\onecolumngrid 
\clearpage


\section*{Supplementary material}

## Supplementary Note I: Coincidence windows, accidentals, and histograms

Fig. S4 shows the unheralded and heralded photon pair coincidence histograms of the measurements with $\ell_{p,s,i} = 0$ presented in Fig. 2 in the main text. The coincidences are summed over the whole measurement period, in total over 18.6 h. The time delay of the heralded photon pair histogram corresponds to the delay between a photon pair and the heralding photon. Note, that the coincidence window of the photon pair coincidence has to be chosen before the measurement, and, to prevent the loss of genuine photon pairs, is chosen to be 1 ns long. The coincidence window of the photon pairs and the heralding photons is chosen in post-processing to be 300 ps wide and is highlighted in light blue. The accidental rate is estimated to be the mean heralded photon rate outside of the coincidence window (highlighted in orange). The coincidence rate of $1.34 \pm 0.39$ heralded photon pairs per hour (corrected for accidentals) is well above the accidental rate of 0.14 per hour per coincidence window. We measure up to $40.18 \pm 1.74$ unheralded photon pair coincidences per hour in the $\ell_{p,s,i} = 0$ projection in a coincidence window of 400 ps.

## Supplementary Note II: Photon statistics of the quantum state used to pump the SPDC

In the presented experiment, we pump an SPDC process with single photons from another SPDC process. Here, we discuss the multi-pair emission by the first SPDC source, to investigate if the generated quantum state is a true single-photon state. For that, we model the first SPDC process induced by a strong, coherent, and undepleted drive field with mean photon number $n_d \gg 1$ by an idealized two-mode squeezed state that takes the form [1]

$$|\psi\rangle_{h,p} = \frac{1}{\cosh \gamma} \sum_{n=0}^{\infty} \tanh^n \gamma \, |n, n\rangle_{h,p}, \quad (S5)$$

where $h$ and $p$ denote the heralding and pump modes, respectively, $n$ the photon number, and $\gamma$ the parametric gain [1, 2] which scales with the length of the crystal, the nonlinear susceptibility, and the strength of the drive field. This state is an approximate solution of the Schrödinger equation using an analogue of the Hamiltonian in the main text where we assume that the drive, heralding, and pump modes are perfectly phase-matched and that the drive field is strong and not depleted. In the low gain regime, where $\gamma \ll 1$, Eq. (S5) can be simplified to

$$|\psi\rangle_{h,p} \approx |0, 0\rangle_{h,p} + \gamma \, |1, 1\rangle_{h,p}. \quad (S6)$$

However, in the experiments with a single-photon pump presented in the main text, the relatively high driving powers might lead to higher-order contributions, i.e. to multiple pairs.

To directly measure the contributions of higher-order terms, we introduce a 50:50 beam splitter in the pump photon arm thereby creating arms $p$ and $p'$. The ratio $R$ between coincidences $C_{pp'}$ that could be related to multi-pair emissions (if there are any) and coincidences stemming from single pairs, i.e., $C_{hp}$ and $C_{hp'}$ would give an estimation of the possible impact of multi-pair emissions on our measurement results. We measure these ratios for 16 different driving power levels from the low gain regime with $p_d = 606\,\mu\text{W}$ to high driving powers with $p_d = 52.7\,\text{mW}$ and find values ranging from $R_{606\mu\text{W}} = 8.0 \pm 5.4 \times 10^{-5}$ to $R_{52.7\text{mW}} = 40 \pm 3 \times 10^{-4}$. Note, that for driving powers of more than 3 mW, we have to include a neutral density filter transmitting only around 10 % in the heralding arm to not exceed the maximally tolerable detection rate of the detectors, which we correct for in the analysis. When we investigate how many of these possible multi-pair detections, i.e., $C_{pp'}$, are caused by genuine multi-pairs by inspecting the respective histograms, we find that essentially all $C_{pp'}$ coincidences are solely caused by accidental coincidence detections. In other words, we find that these ratios are independent of the delay between detectors in arm $p$ and $p'$. Hence, if we correct for the accidental detections, we obtain values distributed around zero leading to an average ratio of all driving powers of $R_{\text{ave}} = 2.6 \times 10^{-6}$ with a standard deviation of $1.6 \times 10^{-3}$. From these results, we conclude that the true contributions of higher-order pair emissions are overshadowed by the accidental rate, which makes it impossible to directly measure them.

To find an estimate for the generation of multi-pairs, we instead find an expression for the interaction parameter $\gamma$ that is scalable with the driving power. From Eq. (S5) we obtain an estimate for the probability $P(n)$ of generating $n$ photon pairs in one coherence time $t_{\text{coh}}$ of the drive field, i.e. $P(n) = \tanh^{2n} \gamma / \cosh^2 \gamma$. The gain parameter

$$\gamma = \alpha_d \kappa. \quad (S7)$$

depends on the square root of the mean photon number of the drive field $\alpha_d = \sqrt{n_d}$ within $t_{\text{coh}}$ and on the effective nonlinear coefficient $\kappa$. This coefficient is to be determined empirically and depends on the characteristics of the nonlinear medium and the involved modes.

In the low-gain regime, we use a driving power $p_d = 614\,\mu\text{W}$. The drive laser has a coherence time of $t_{\text{coh}} = 0.3\,\text{ns}$, taken from the laser specifications, and a wavelength $\lambda_d = 524.59\,\text{nm}$. With $\omega_d = 2\pi c/\lambda_d$, where $c$ is the speed of light,

$$\alpha_d = \sqrt{n_d} = \sqrt{\frac{p_d}{\hbar \omega_d} t_{\text{coh}}} \approx 697 \quad (S8)$$

follows. The measured heralding and pump coincidence rate at this drive power is $C_{h,p} = 216\,\text{kHz}$ with a single

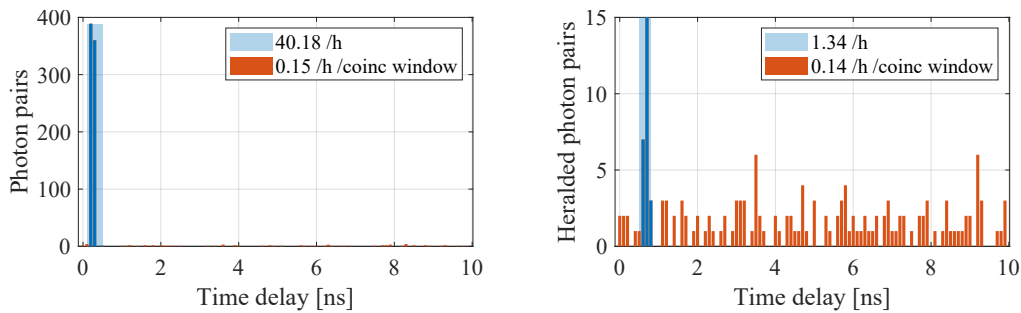

FIG. S4. Histograms of unheralded (left) and heralded (right) photon pairs. The histograms are taken during the measurement presented in Fig. 2 in the main text for the projection on $\ell_{p,s,i} = 0$. The coincidences are summed up over a total measurement time of 18.6 h. The coincidence windows are highlighted in light blue. The accidental rates are estimated by the mean accidental rate outside of the coincidence window.

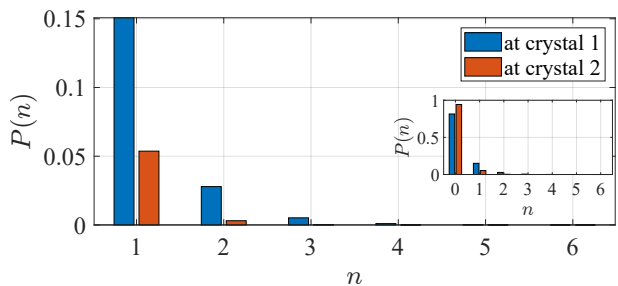

FIG. S5. Estimated photon statistics of the pump mode. The figure shows the probability $P(n)$ of the pump mode having $n$ photons immediately after the first crystal (blue bars) and immediately before the second crystal (orange bars). The inset shows the same plot starting from $n = 0$ to display the vacuum contribution.

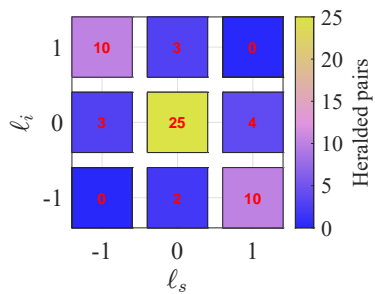

FIG. S6. Number of counts for the measurements presented in Fig. 2 in the main text.

pump rate of $S_p = 1.13$ MHz. With the coupling efficiency $\eta_{\mathrm{coup}} = C_{h,p}/S_p$, the measured pair generation probability per coherence time is

$$P(1) = \frac{S_p}{\eta_{\mathrm{coup}}} t_{\mathrm{coh}} \approx 1.8 \times 10^{-3}. \tag{S9}$$

Together with $P(1) = \gamma^2$, which we obtain from Eq. (S6), we can now find the parametric gain factor $\kappa = \gamma/\alpha_d \approx 6.07 \times 10^{-5}$.

Finally, in the high gain regime, the interaction parameter $\gamma$ is now estimated by adjusting $\alpha_d$ with Eq. (S8)

with the desired drive power. The result obtained for $p_d = 72.7$ mW is shown in Fig. S5 in blue bars, where we have used $P(n) = \tanh^{2n} \gamma / \cosh^2 \gamma$.

To further estimate the pump photon state including the losses $\eta$ suffered before reaching the second SPDC source, we study the SMF coupling and the diffraction efficiency of $\mathrm{SLM_p}$. We estimate the detection efficiency of the single-photon avalanche diodes measuring the pump photons to be on the order of $\eta_{\mathrm{det}} = 50\%$, such that the losses due to the SMF coupling are $\eta_{\mathrm{SMF}} = \eta_{\mathrm{coup}}/\eta_{\mathrm{det}} = 38\%$. With $\eta_{\mathrm{SLM}} = 0.7$, we arrive at the overall efficiency $\eta = \eta_{\mathrm{SMF}}\eta_{\mathrm{SLM}} = 27\%$. We include these efficiencies by transforming the probability distribution $P(n)$ with

$$P(n) \rightarrow \sum_{j=n}^{\infty} P(j)\eta^n (1-\eta)^{j-n} \binom{j}{n}, \tag{S10}$$

which can be derived by modelling the losses as a beam-splitting operation [3]. The result is shown in Fig. S5 in red bars. From our estimates, we retrieve a ratio $P(1)/P(> 1) = 16.56$ at the second crystal, i.e., for roughly every 16 down-conversion events one event does not come from a single-photon pump.

## Supplementary Note III: Total number of counts

For reference, Fig. S6 and S7 show the total number of coincidence counts for all projections of the measurements presented in the main text.

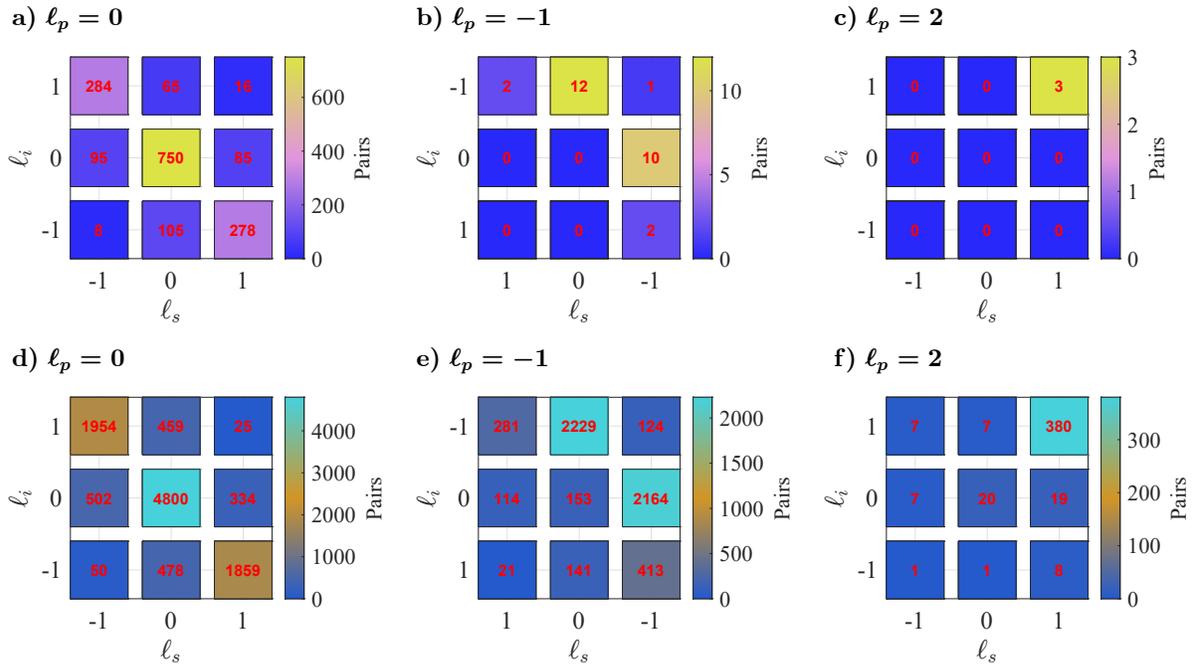

FIG. S7. Number of counts for all measurements presented in Fig. 3 in the main text.



\end{document}